\documentclass[aps,prl,preprint,superscriptaddress,preprintnumbers,byrevtex,amsmath,amssymb,showpacs,byrevtex]{revtex4}

\usepackage{graphicx} 
\usepackage{subfigure}
\usepackage{epsfig}

\renewcommand{\arraystretch}{1.1}


\begin{document}
\epsfysize 25mm
\epsfbox{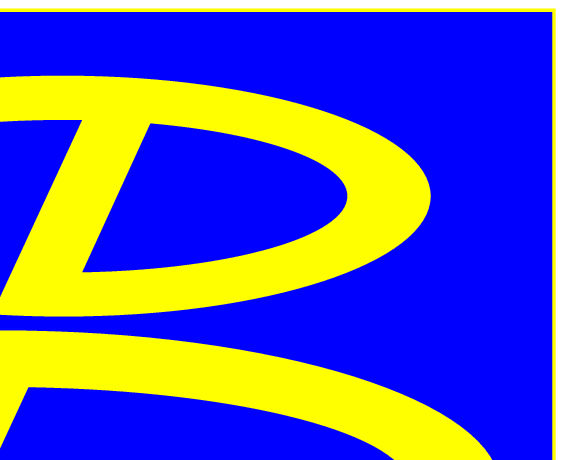}
\begin{flushright}
\vskip -25mm
\noindent
\hspace*{3.0in}{\bf Belle preprint 2003-7} \  \\
\hspace*{3.0in}{\bf KEK preprint 2003-24} \ 
\end{flushright}
\vskip 25mm

\begin{center}
 \quad\\[1cm] \Large \boldmath \bf
{ Comment on ``$e^+ e^{-}$ annihilation into $J/\psi +
J/\psi$''} 
\end{center}


{\renewcommand{\thefootnote}{\fnsymbol{footnote}}
\normalsize
\tighten

\begin{center}
  K.~Abe$^{10}$,              
  K.~Abe$^{43}$,              
  R.~Abe$^{30}$,              
  T.~Abe$^{44}$,              
  I.~Adachi$^{10}$,           
  Byoung~Sup~Ahn$^{17}$,      
  H.~Aihara$^{45}$,           
  M.~Akatsu$^{23}$,           
  Y.~Asano$^{50}$,            
  T.~Aso$^{49}$
% ...
  Y.~Yuan$^{12}$,             
  Y.~Yusa$^{44}$,             
  Z.~P.~Zhang$^{37}$,         
  V.~Zhilich$^{2}$,           
and
  D.~\v Zontar$^{50}$         
\end{center}

\begin{center}
  (Belle Collaboration)
\end{center}

\renewcommand{\baselinestretch}{1.0} \small
\begin{center}
% ...
$^{2}${Budker Institute of Nuclear Physics, Novosibirsk}\\
% ...
$^{10}${High Energy Accelerator Research Organization (KEK), Tsukuba}\\
% ...
$^{12}${Institute of High Energy Physics, Chinese Academy of Sciences, 
Beijing}\\
% ...
$^{17}${Korea University, Seoul}\\
% ...
$^{23}${Nagoya University, Nagoya}\\
% ...
$^{30}${Niigata University, Niigata}\\
% ...
$^{37}${University of Science and Technology of China, Hefei}\\
% ...
$^{43}${Tohoku Gakuin University, Tagajo}\\
$^{44}${Tohoku University, Sendai}\\
$^{45}${University of Tokyo, Tokyo}\\
% ...
$^{49}${Toyama National College of Maritime Technology, Toyama}\\
$^{50}${University of Tsukuba, Tsukuba}\\
% ...
\end{center}

\newpage

\renewcommand{\arraystretch}{1.1}
\renewcommand{\baselinestretch}{1.2} \normalsize
\setcounter{footnote}{0}

\noindent
The first observations of annihilation processes of the type $e^+
e^-\to J/\psi\, \eta_c$ and $J/\psi\, (c\bar{c})_{\text{non-res}}$
were recently reported by Belle~\cite{2cc}.  The measured
cross-sections for both processes are an order-of-magnitude larger
than theoretical predictions based on non-relativistic QCD
(NRQCD)~\cite{pred-psi-eta,pred-psi-ccbar}.  In an attempt to explain
at least part of this discrepancy, the authors of Ref.~\cite{bra_psi}
suggest that processes proceeding via two virtual photons may be
important.  In particular, if the two-photon-mediated process
$e^+e^-\to J/\psi\, J/\psi$ has a significant cross-section, the
observed $e^+ e^- \to J/\psi\, \eta_c$ signal, which is inferred from
the $\eta_c$ peak in the recoil mass spectrum for the reconstructed
$J/\psi$ in inclusive $e^+ e^- \to J/\psi\, X$ events, might also
include double $J/\psi$ events and, thus, produce an inflated
cross-section measurement.  $e^+e^-$ annihilation to $J/\psi\, J/\psi$
via a single virtual photon is forbidden by charge conjugation
symmetry and was ignored in our published analysis.  Here, using a
data sample of 101.8 fb$^{-1}$ collected by the Belle detector and the
analysis procedure described in Ref.~\cite{2cc}, we evaluate this
possibility.

Since the $\eta_c$ and $J/\psi$ have similar masses
($M_{J/\psi}-M_{\eta_c} \simeq 116$~MeV/$c^2$),  it is
important to check for any momentum scale bias that may
shift the recoil mass values.  We use $e^+ e^-\to \psi(2S)\, \gamma$,
$\psi(2S)\to J/\psi\, \pi^+\pi^-$ events to calibrate and verify the
recoil mass scale.  We find that any shift in the recoil mass is less
than 3~MeV/c$^2$.

The spectrum of recoil masses against the $J/\psi$ 
is presented in Fig.~\ref{cc3}: a clear peak is observed around the $\eta_c$
nominal mass, and a smaller peak is seen around the $\chi_{c0}$ nominal mass;
the large peak at $\sim 3.63\,\mathrm{GeV}/c^2$
is interpreted as the $\eta_c(2S)$.
We performed a fit to this spectrum that includes
all of the known narrow charmonium states. In this
fit, the mass positions for the $\eta_c$, $\chi_{c0}$ and $\eta_c(2S)$
are treated as free parameters; those for the $J/\psi$, $\chi_{c1}$, $\chi_{c2}$
and $\psi(2S)$ are fixed at their nominal values.
The expected line-shapes for these peaks are
determined from a Monte Carlo simulation as described in our
previous paper~\cite{2cc},
the background is parametrized by a second order polynomial
function, and only the region below the open charm threshold
($M_{\rm recoil} < 3.7 \, \mathrm{GeV} /c^2$) is included in the fit.
The fit results, listed in Table~\ref{cc3t}, give
negative yields for the $J/\psi$, $\chi_{c1}$, $\chi_{c2}$ and
$\psi(2S)$;  the solid line in Fig.~\ref{cc3} is
the result of a fit with all these contributions fixed at zero.
The dotted line in the figure corresponds to the case where the
contributions of the $J/\psi$, $\chi_{c1}$, $\chi_{c2}$ and $\psi(2S)$
are set at their 90$\%$ confidence level upper limit values.  The dashed
line is the background function.  To set a conservative upper limit
for $e^+ e^- \to J/\psi\, J/\psi$, we use assumptions for
the production and helicity angle distributions that
correspond to the lowest detection efficiency.

\begin{figure}[ht]
\hspace*{-0.05\textwidth} \includegraphics[width=0.8\textwidth,height=0.48\textwidth,bb=0 20 568 341]
{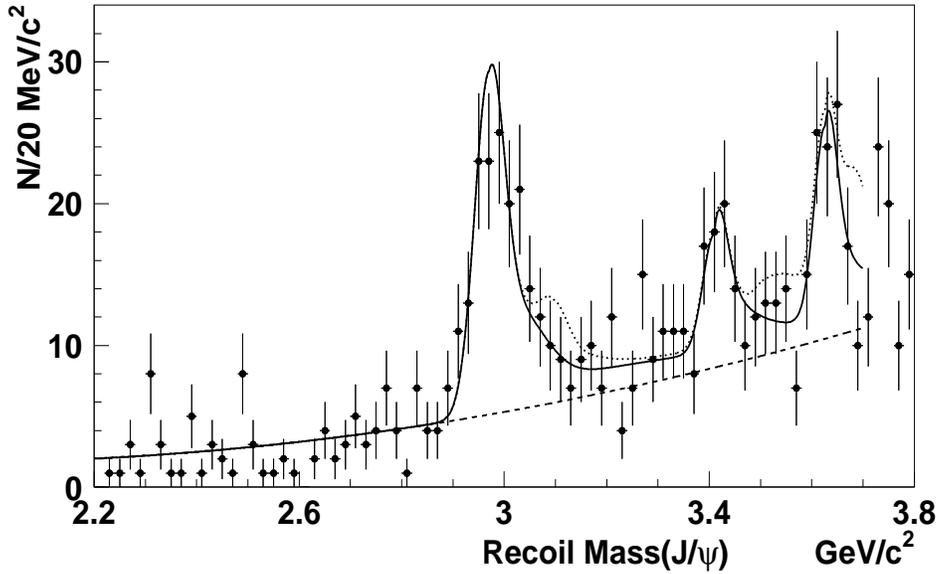}
\caption{Distribution of masses recoiling against the reconstructed
$J/\psi$ in inclusive $e^+e^-\to J/\psi\, X$ events. The
curves are explained in the text.}
\label{cc3}
\end{figure}

\begin{table}[ht]
  \caption{Summary of the signal yields, charmonium masses and significances
	for $e^+ e^- \to J/\psi \, (c\bar{c})_{\text{res}}$.}
  \label{cc3t}
  \begin{ruledtabular}
    \begin{tabular}{lrcc}
	$(c\bar{c})_{\text{res}}$ state
			& \multicolumn{1}{c}{$N$}
					& $M\,[\mathrm{GeV}/c^2]$
								& $\sigma$
									\\\hline 
	$\eta_c$	& $175 \pm 23$	& $2.972 \pm 0.007$	& $9.9$	\\
	$J/\psi$	& $-9  \pm 17$	& fixed			& ---	\\
	$\chi_{c0}$	& $61  \pm 21$	& $3.409 \pm 0.010$	& $2.9$	\\
  $\chi_{c1}+\chi_{c2}$	& $-15 \pm 19$	& fixed			&  ---	\\
	$\eta_c(2S)$	& $107 \pm 24$	& $3.630 \pm 0.008 $	& $4.4$	\\
	$\psi(2S)$	& $-38 \pm 21$	& fixed			& ---	\\
    \end{tabular}
  \end{ruledtabular}
\end{table}

In summary, using a larger data set we confirm our
published observation of
$e^+ e^- \to J/\psi\, \eta_c$ and find no evidence for the process
$e^+ e^- \to J/\psi\, J/\psi$.  We set an upper limit for
$\sigma(e^+ e^- \to J/\psi\, J/\psi) \times
\mathcal{B}(J/\psi \to \, >2\,\text{charged})$ of less than $0.008 \,
\mathrm{pb}$ at the $90\%$ CL.

Although the limit presented here is not inconsistent with the
prediction for the $J/\psi\,J/\psi$ production rate given
in Ref.~\cite{bra_psi}, the suggestion that a significant fraction of the
inferred $J/\psi\,\eta_c$ signal is actually $J/\psi\,J/\psi$
is ruled out.  Therefore, the discrepancy between the Belle result
and the NRQCD prediction remains.

\normalsize



\end{document}